\pdfoutput=1
\documentclass[sigplan, twocolumn]{acmart}

\usepackage{xurl}
\usepackage{hyperref}
\usepackage{amsmath}
\usepackage{xspace}
\usepackage{multirow}
\usepackage{placeins}
\usepackage{algorithm}
\usepackage{algpseudocode}
\algrenewcommand\algorithmicrequire{\textbf{Input:}}
\algrenewcommand\algorithmicensure{\textbf{Output:}}
\algrenewcommand{\algorithmiccomment}[1]{\hfill\textcolor{gray}{$\triangleright$~#1}}
\usepackage{subcaption}
\graphicspath{{./}}

\newcommand{\SystemName}{MpFA\xspace}

\renewcommand\footnotetextcopyrightpermission[1]{}

\setcopyright{none}
\copyrightyear{2027}
\acmYear{2027}
\acmDOI{}
\acmConference[EuroSys '27]{The European Conference on Computer Systems}{April 2027}{Morocco}
\acmISBN{}

\title{MpFA: Hardware-Efficient Train-Free QK4V8 FlashAttention Kernels on Blackwell GPUs}

\author{
  Chencheng Deng, Jianbin Fang, Dezun Dong\textsuperscript{*}
}
\affiliation{%
  \institution{College of Computer Science and Technology, National University of Defense Technology}
  \city{Changsha}
  \country{China}
}
\email{{chenchengdeng, j.fang, dong}@nudt.edu.cn}

\begin{document}

\makeatletter
\fancypagestyle{standardpagestyle}{%
  \fancyhf{}%
  \renewcommand{\headrulewidth}{\z@}%
  \renewcommand{\footrulewidth}{\z@}%
  \fancyfoot[C]{\if@ACM@printfolios\footnotesize\thepage\fi}%
}
\pagestyle{standardpagestyle}
\makeatother

\begin{abstract}
Long-context LLM inference pushes modern GPU serving stacks into an attention-bound regime, 
where both compute and memory are dominated by the softmax--GEMM pipeline. 
On NVIDIA Blackwell GPUs, FP4 Tensor Cores offer high matmul throughput, 
but we find that fully FP4 attention often fails to translate this throughput into end-to-end speedups due to non-matmul costs: online quantization after softmax, tensor/shared-memory data movement, and contention on the softmax path.
We present \SystemName, a training-free FlashAttention kernel optimized for Blackwell. 
Guided by hardware characterization, \SystemName uses mixed precision: NVFP4 for $QK^\top$ and FP8 for $PV$ (QK4PV8). This preserves low-bit $QK^\top$ throughput while avoiding the conversion and scaling overheads of FP4 $PV$. To recover accuracy without further stressing the softmax pipeline, \SystemName introduces rank-one smoothing compensation implemented as an additional Tensor Core MMA.
\SystemName further improves performance with a fine-grained asynchronous pipeline, 
tensor-memory reuse, and adaptive parallel partitioning across prefill and decode. 
On an NVIDIA SM100 GPU and across $16$K--$128$K contexts, \SystemName improves prefill 
throughput over state-of-the-art BF16/FP8 baselines and increases end-to-end output 
throughput by $2.81\times$ over BF16 FA4 across Llama-3.1-8B and Qwen3-14B. 
Across five benchmark suites and two models, rank-one compensation 
recovers $62.5\%$ of the accuracy loss with about $2.0\%$ kernel overhead.
\end{abstract}


\maketitle

\renewcommand{\thefootnote}{}%
\footnotetext{\textsuperscript{*}Corresponding author.}%
\renewcommand{\thefootnote}{\arabic{footnote}}

\section{Introduction}
\label{sec:intro}

Applications such as retrieval-augmented generation, code understanding, and
multi-turn agents increasingly rely on long contexts, making long-context
inference an important LLM-serving workload~\cite{longbench,ruler}. Meanwhile,
AI infrastructure is adopting the Blackwell architecture. NVIDIA's data-center
SM100 GPUs combine fifth-generation Tensor Cores, native low-bit computation, and
tensor memory for storing operands and intermediate results.

\begin{figure}[t]
\centering
\includegraphics[scale=1]{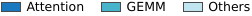}\\[-1pt]
\begin{subfigure}[t]{0.49\columnwidth}
  \centering\includegraphics[width=\linewidth]{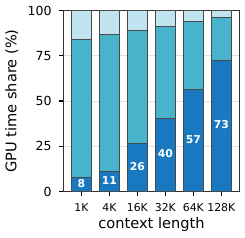}
  \caption{prefill}\label{fig:share:pre}
\end{subfigure}\hfill
\begin{subfigure}[t]{0.49\columnwidth}
  \centering\includegraphics[width=\linewidth]{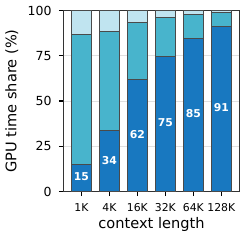}
  \caption{decode}\label{fig:share:dec}
\end{subfigure}
\caption{Device-side GPU time breakdown for Llama-3.1-8B served by SGLang
on an SM100 GPU with the BF16 FA4 backend.}
\vspace{0mm}
\label{fig:share}
\end{figure}

LLM inference consists of prefill and decode. Prefill processes the entire
input sequence, so its attention computation grows approximately quadratically with
context length. Decode generates one token at a time and repeatedly accesses
the growing KV cache, resulting in memory traffic that increases linearly with
context length and low arithmetic intensity. We run Llama-3.1-8B with SGLang's standard BF16 FA4
backend on an SM100 GPU. As shown in Figure~\ref{fig:share}, attention accounts for
$72.5\%$ of device-side GPU time in prefill and $91.4\%$ in decode at $128$K
tokens. Although FlashAttention-4 (FA4)~\cite{fa4} pipelines computation through
tensor memory and fifth-generation Tensor Cores, the evaluated backend uses BF16 and does not exploit
FP4 Tensor Cores for the two dominant matrix multiplications.

Quantization-aware training can adapt a model to 4-bit attention, but requires
costly per-model retraining and is difficult to deploy at scale.
Post-training methods avoid retraining; for example, SageAttention3~\cite{sageattn3}
applies NVFP4 attention to consumer Blackwell SM120 GPUs (RTX~5090), while
HiFA4~\cite{hifa4} targets the Ascend HIF4 NPU. However, neither is designed
to exploit SM100's tensor memory, fifth-generation Tensor Cores, and
operand-delivery paths or to jointly support decode optimization, paged KV
caches, and CUDA Graph capture. Efficient training-free FP4
attention therefore remains unavailable on data-center Blackwell GPUs.

Our hardware characterization and profiling reveal two challenges in building
training-free FP4 FlashAttention for data-center Blackwell GPUs. First, when 
fifth-generation Tensor Cores execute block-scaled FP4
matrix multiply-add (MMA), matrix operands must be supplied from shared
memory, whereas scale factors must come from tensor memory. Quantizing the
softmax output $P$ to FP4 adds per-block amax computation, scale-factor
generation, and data movement between tensor and shared memory. These
operations load the softmax path, preventing low-bit matrix-multiplication
speedups from translating into end-to-end
performance. Second, existing smoothing compensation methods place the
compensation term on the softmax path. This extra element-wise work further
loads the already saturated path, where it competes with other operations and
offsets the benefit of FP4 computation.

We present \SystemName, a high-performance, training-free low-bit attention
kernel for LLM inference on data-center Blackwell GPUs. We characterize three
precision assignments: NVFP4 for both matrix multiplications; NVFP4 for
$QK^\top$ with BF16 for $PV$; and NVFP4 for $QK^\top$ with FP8 for $PV$.
Our evaluation shows that QK4PV8 provides the best balance between arithmetic
throughput and non-matmul overhead on SM100, making it the
hardware-efficient choice for FA4. FP8 $PV$ retains the benefits of
low-precision Tensor Core execution while avoiding the additional
precision-conversion requirements of an FP4 probability matrix. BF16 $PV$
avoids these conversion requirements as well, but leaves Tensor Core throughput
on the table because FP8 Tensor Cores provide twice the throughput of BF16.
\SystemName therefore adopts QK4PV8 mixed precision and optimizes the FA4
pipeline for both inference stages: prefill uses a fine-grained asynchronous
pipeline and tensor-memory column reuse to co-locate scale factors with
double-buffered accumulators; decode restores parallelism through GQA packing,
KV partitioning, and column-split softmax while avoiding exponential
evaluations on padded rows.

To reduce smoothing overhead, \SystemName introduces a rank-one Tensor Core
MMA compensation method that moves the correction from the saturated softmax
path to otherwise underutilized Tensor Cores. It stores $K$ in
NVFP4 and $V$ in FP8, consuming the low-bit KV cache without dequantization
and reducing KV bytes per token to $1/2.56$ of BF16. 

Experiments on SM100 show that, across $16$K--$128$K, \SystemName achieves
average prefill speedups of $1.31$--$1.33\times$ over BF16 FA4 and
$1.15$--$1.17\times$ over FP8 FlashInfer. Across $8$K--$128$K, its
decode kernel achieves a $2.08\times$ average speedup over split-KV-enabled
BF16 FA4. Across ten model--task pairs spanning five standard benchmarks,
rank-one smoothing recovers $62.5\%$ of the accuracy loss on average, with
about $2.0\%$ average kernel overhead. 
\SystemName achieves $2.81\times$ the end-to-end output throughput of BF16
FA4 across both models and $16$K--$128$K, and by $1.06\times$ over FP8
FlashInfer across both models and $32$K--$128$K contexts. 

This paper makes the following contributions:
\vspace{-\listisep}
\begin{enumerate}
\item We systematically analyze how attention kernels on data-center
Blackwell GPUs become limited by non-matmul units, providing
insight for optimizing other fused operators (\S\ref{sec:motivation}).
\item We present \SystemName, a high-performance QK4PV8 Flash-Attention kernel
for LLM inference, with tensor memory reuse and stage-specific optimizations
for prefill and decode (\S\ref{sec:kernel}).
\item We introduce a rank-one Tensor Core smoothing algorithm that moves
compensation from the saturated softmax path to MMA, recovering accuracy while
maintaining hardware utilization (\S\ref{sec:quant}).
\end{enumerate}

\section{Background}
\label{sec:background}

\subsection{Blackwell GPUs}
\label{sec:back:hw}

\textbf{Blackwell} is NVIDIA's GPU architecture for next-generation AI
computing. This paper focuses on data-center SM100 rather than
consumer SM120 (RTX~5090). Both provide Blackwell Tensor Cores and low-bit formats,
but differ in compute capabilities and operand delivery. SM120 attention
kernels are optimized for the consumer GPU's matrix multiply-add (MMA) path,
whereas SM100 introduces on-chip tensor memory and
\texttt{tcgen05.mma}. Consequently, SM120 implementations cannot fully
exploit SM100, whose adoption requires attention kernels redesigned around
its data path~\cite{blackwellbench}.

\begin{figure}[t]
\centering
\begin{subfigure}[t]{1\columnwidth}
\centering
\includegraphics[width=\linewidth]{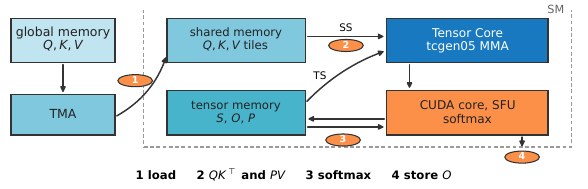}
\caption{FlashAttention4 data path}
\label{fig:hw:path}
\end{subfigure}

\begin{subfigure}[t]{\columnwidth}
\centering
\includegraphics[width=0.8\linewidth]{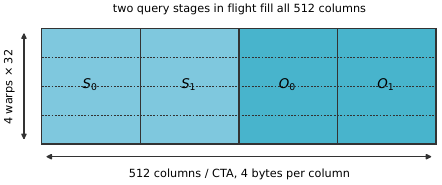}
\caption{Tensor memory architecture}
\label{fig:hw:tmem}
\end{subfigure}
\caption{Data path and tensor memory organization.}
\vspace{0mm}
\label{fig:hw}
\end{figure}

\noindent\textbf{From hardware to kernels.}
SM100's Tensor Cores, tensor memory, and shared memory jointly determine
the on-chip data path for attention. Figure~\ref{fig:hw} summarizes the
execution path considered in this paper. In Figure~\ref{fig:hw}(a), the
asynchronous copy engine (TMA) first moves tiled $Q$, $K$, and $V$ from
global memory to shared memory. Tensor Cores then use \texttt{tcgen05.mma} instructions 
to compute $QK^\top$ and
$PV$ and write the intermediate results to tensor memory. CUDA cores and
special function units execute online softmax and directly read from and
write to the score, probability, and output accumulators in tensor memory.
The final results are written back to global memory. The SS and TS labels in
the figure denote two operand-delivery paths; as discussed in
\S\ref{sec:motiv:gap}, these paths constrain the precision assignment.
Unlike traditional execution models, which keep accumulators in registers,
this design uses tensor memory as separate on-chip storage for results,
allowing matrix multiplication and softmax to
form an asynchronous pipeline with explicit role separation~\cite{fa4}.
Figure~\ref{fig:hw}(b) shows the capacity constraint: in our kernel
configuration, each cooperative thread array (CTA) is allocated $128$ rows $\times$ $512$ columns, or
$256$ KB. Two attention score accumulators and two output accumulators, each with
block size $128$ and FP32 elements, consume all $512$ columns.

\noindent\textbf{Block-scaled FP4 matrix multiply-add.}
FP4 cannot represent the full dynamic range of activations by itself, so
SM100 uses block-scaled FP4 MMA. NVFP4 associates every consecutive
$G{=}16$ elements along the reduction dimension with one \texttt{e4m3}
scale factor; by comparison, MXFP4 uses $G{=}32$ elements and one
\texttt{e8m0} scale factor. For a block $x_g\in\mathbb{R}^{G}$, quantization
and hardware reconstruction can be written as
\begin{equation}
s_g = \frac{\mathrm{amax}(x_g)}{\rho},\qquad
\hat{x}_g = \Big\lfloor \frac{x_g}{s_g} \Big\rceil{\mathrm{e2m1}},\qquad
\tilde{x}_g = s_g\hat{x}_g,
\label{eq:nvfp4}
\end{equation}
where $\rho{=}6$ is the maximum magnitude of \texttt{e2m1}. In
Equation~\ref{eq:nvfp4}, the Tensor Core performs the
reconstruction $\tilde{x}_g=s_g\hat{x}_g$ internally during MMA.
Block scaling limits quantization error locally, but introduces scale
factors that must accompany the operands. We use NVFP4 because its
$16$-element blocks and scales better match attention
activations than MXFP4~\cite{sageattn3}. Different MMA precisions impose
different operand-source and scale-factor requirements, making hardware
details critical to both precision selection and performance bottlenecks.

\subsection{FlashAttention and Quantization}
\label{sec:back:fa}

Self-attention comprises two matrix multiplications and one softmax:
\begin{equation}
S=QK^\top/\sqrt{d},\qquad P=\mathrm{softmax}(S),\qquad O=PV
\end{equation}
FlashAttention~\cite{fa1,fa2} tiles $Q$, $K$, and $V$ along the sequence
dimension and processes KV blocks with online softmax, keeping intermediates
on chip. Let the score for the $t$-th KV block be
$S^{(t)}=QK^{(t)\top}/\sqrt{d}$. FlashAttention maintains a row-wise maximum
$m$, normalization factor $\ell$, and output accumulator $O$:
\begin{align}
m^{(t)} &= \max\!\big(m^{(t-1)},\,\mathrm{rowmax}(S^{(t)})\big)
\label{eq:osmax}\\
P^{(t)} &= e^{S^{(t)}-m^{(t)}}
\label{eq:osexp}\\
\ell^{(t)} &= e^{\,m^{(t-1)}-m^{(t)}}\ell^{(t-1)}
  + \mathrm{rowsum}\big(P^{(t)}\big)
\label{eq:ossum}\\
O^{(t)} &= \mathrm{diag}\!\big(e^{\,m^{(t-1)}-m^{(t)}}\big)O^{(t-1)}
  + P^{(t)}V^{(t)}
\label{eq:osout}
\end{align}
Row maximum, exponentiation, row sum, and output rescaling in
Equations~\ref{eq:osmax}--\ref{eq:osout} form the non-matmul softmax path.
FA3~\cite{fa3} hides its latency through warp specialization, overlapping a
tile's matmul with the preceding tile's softmax. FA4~\cite{fa4} reorganizes
this pipeline for Blackwell tensor memory and fifth-generation Tensor Cores
(Figure~\ref{fig:hw}(a)).

\noindent\textbf{Quantized attention.}
Unlike static weights in linear-layer quantization, $Q$, $K$, $P$, and $V$
are runtime activations coupled by softmax, making quantized attention
sensitive to outliers, normalization, and accumulation error. The SageAttention
series~\cite{sageattn,sageattn2,sageattn3} and HiFA4~\cite{hifa4} demonstrate
that post-training quantization can accelerate attention without retraining.
It can also consume low-bit KV caches directly~\cite{kivi}, reducing memory
use and avoiding dequantization.

\section{Motivation}
\label{sec:motivation}
\begin{table*}
\centering
\small
\setlength{\tabcolsep}{7pt}
\renewcommand{\arraystretch}{1.25}
\caption{Positioning relative to the closest related systems.}
\label{tab:pos}
\begin{tabular}{l|lllll}
\hline
Work & Hardware & $QK^\top$ / $PV$ & Compensation & Partitioned decode & Workload \\
\hline
FA4~\cite{fa4} & SM100 & BF16 / BF16 & --- & Incompatible with graph capture & Vision/language \\
FlashInfer~\cite{flashinfer} & SM100 & FP8 / FP8 & --- & Graph-compatible & Vision/language \\
SageAttn~\cite{sageattn} & SM89 & INT8 / INT8 & Softmax & --- & Vision/language \\
SageAttn2~\cite{sageattn2} & SM90 & INT4 / FP8 & Softmax & --- & Vision/language \\
SageAttn3~\cite{sageattn3} & SM120 & NVFP4 / NVFP4 & Softmax & --- & Vision \\
HiFA4~\cite{hifa4} & Ascend NPU & HIF4 / HIF4 & Softmax & --- & Language \\
\textbf{\SystemName (ours)} & \textbf{SM100} & \textbf{NVFP4 / FP8} &
\textbf{MMA} & \textbf{Graph-compatible} & \textbf{Language} \\
\hline
\end{tabular}
\end{table*}

\begin{figure*}[t]
\centering
\includegraphics[scale=1]{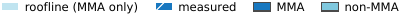}\\[-1pt]
\begin{subfigure}[t]{0.315\textwidth}
  \centering\includegraphics[width=\linewidth]{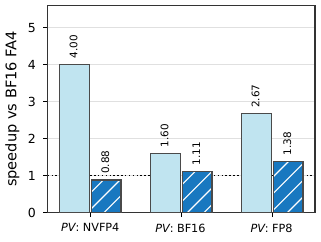}
  \caption{$QK^\top$ fixed to NVFP4}\label{fig:motiv:a}
\end{subfigure}\hfill
\begin{subfigure}[t]{0.335\textwidth}
  \centering\includegraphics[width=\linewidth]{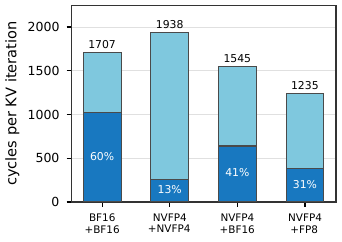}
  \caption{Performance gap}\label{fig:motiv:b}
\end{subfigure}\hfill
\begin{subfigure}[t]{0.335\textwidth}
  \centering\includegraphics[width=\linewidth]{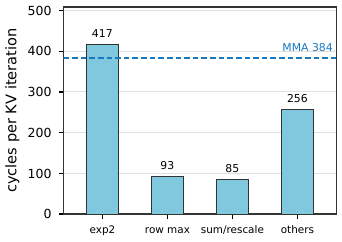}
  \caption{Performance breakdown}\label{fig:motiv:c}
\end{subfigure}
\caption{Four precision assignments for causal prefill at $L{=}64$K with
GQA $32{:}8$.}
\vspace{2mm}
\label{fig:motiv}
\end{figure*}

Low-bit attention can exploit SM100's FP4 Tensor Cores while compressing
the KV cache, reducing the compute and memory costs of long-context inference.
However, Table~\ref{tab:pos} shows that existing systems target consumer
Blackwell GPUs or NPUs, without addressing SM100's tensor memory,
operand-delivery paths, or serving-time decode.

\subsection{Asymmetric Hardware Scaling}
\label{sec:motiv:gap}

Blackwell's gains concentrate in Tensor Cores: SM100 delivers
$2.25$\,PFLOPS in BF16, compared with $1$\,PFLOPS on SM90, and
$9$\,PFLOPS in FP4~\cite{blackwellbrief,h100arch}. Exponential throughput and
shared-memory bandwidth have not scaled
proportionally~\cite{blackwellbench,cuda124,fa4}, so reducing matmul precision yields no
comparable reduction in softmax cost.
The SM100 operand-delivery paths further amplify this asymmetry. 
Block-scaled FP4 MMA requires its operands to reside in shared memory and 
its scale factors to reside in tensor memory, whereas FP8 MMA can source its first 
operand directly from tensor memory.
Moreover, BF16 FA4 double buffers the queries, while its two FP32 score and output 
accumulators, each with a tile size of $128$, already occupy all $512$ tensor memory columns 
available to each CTA, leaving limited capacity for additional scale factors. 
Precision selection is therefore jointly constrained by compute throughput, data movement, 
and contention for tensor memory capacity.

We implement and measure four configurations on SM100: BF16 FA4, and
three configurations that fix $QK^\top$ to NVFP4 while using NVFP4, BF16,
or FP8 for $PV$. Based only on Tensor Core MMA cycles,
NVFP4 $PV$ should be fastest, followed by FP8 and then BF16, with theoretical
speedups of $4.00\times$, $2.67\times$, and $1.60\times$, respectively.
However, Figure~\ref{fig:motiv}(a) shows the opposite measured ordering.
At $L{=}64$K, NVFP4 $+$ FP8 is $1.38\times$ faster than BF16 FA4,
NVFP4 $+$ BF16 achieves $1.11\times$, while NVFP4 $+$ NVFP4 reaches only
$0.87\times$ and is slower than BF16 FA4. The result demonstrates that precision
selection on SM100 cannot be based on Tensor Core peak throughput alone.

Figure~\ref{fig:motiv}(b) decomposes the SM cycles per KV-block iteration
and explains why the theoretical matrix-multiplication speedup does not
translate into end-to-end operator speedup. In the BF16 configuration, the
two matrix multiplications consume $1024$ cycles, or $60\%$ of total operator
time. The BF16 FA4 implementation already approximates the exponential with a
third-order polynomial on the CUDA cores to reduce the cost of the softmax
exponential, and we adopt the same technique.
NVFP4 $+$ FP8 reduces this component to $384$ cycles, a $63\%$
reduction, but total time falls only from $1707$ to $1235$ cycles, or $28\%$.
Softmax, reductions, and data movement do not shrink with matrix-multiplication
precision. Once attention enters the 4-bit regime, the dominant constraint
therefore shifts from matrix multiplication to the softmax path.

The reason NVFP4 $+$ NVFP4 becomes slower is operand delivery rather than FP4
MMA throughput. If the probability matrix $P$, just generated in tensor
memory by softmax, is used as an operand for block-scaled NVFP4 MMA, three
additional costs arise. First, $P$ must be quantized by computing an
absolute maximum over every $16$ elements and generating scale factors; this
work is placed on the already busy softmax path. Second, block-scaled MMA
requires both operands from shared memory, so $P$ must leave tensor memory,
be written to shared memory, and then be read by the matrix multiplication
unit. Third, $P$'s scale factors need tensor memory capacity, competing with
the score and output accumulators that already consume all $512$ columns and
requiring additional scale-factor transfers.

The FP8 $PV$ path avoids all three costs. Without block scaling, FP8 MMA can
read its first operand directly from tensor memory, allowing the probability
matrix to remain in place. FP8 Tensor Cores also provide twice the throughput
of BF16. Thus, an efficient SM100 precision assignment must consider both data
format and operand-delivery path: NVFP4$(QK^\top)+$FP8$(PV)$ is the
hardware-efficient point.

\subsection{Inefficient Accuracy Compensation}
\label{sec:motiv:comp}

Figure~\ref{fig:motiv}(c) further breaks down the execution time of
NVFP4 $+$ FP8. We obtain the softmax-stage costs through a compile-time
ablation chain: we first rebuild the kernel with \texttt{exp2} disabled, then
additionally disable the row-max and row-sum/rescale stages, and compute
successive differences. This procedure makes the stage costs additive.

The softmax path consumes $595$ cycles, or $48\%$ of total operator time, 
with \texttt{exp2} alone accounting for $417$ cycles, exceeding the $384$ cycles 
of the two quantized matrix multiplications. An $M{=}N{=}128$ KV block 
contains $16384$ score elements. At $16$ elements per SM cycle, 
fully serialized exponential evaluation would require $1024$ cycles. 
The measured $417$ cycles are lower because CUDA core polynomial evaluation 
overlaps with exponential unit execution, while matrix multiplication overlaps with the softmax path. Thus, once attention enters the 4-bit regime, non-matmul computation, particularly the softmax path, becomes the dominant bottleneck.

Low-bit quantization still requires accuracy compensation. Prior methods
typically smooth outliers by subtracting the means of $Q$ and $K$ and adding a
compensation factor back to the attention scores. The compensation must be
added before masking and before updating the running maximum; otherwise it
breaks online-softmax normalization. Existing approaches consequently place
additional loads and element-wise additions on the softmax bottleneck.
In our measurements, this placement costs $18.6$--$22.7\%$ of operator time,
consuming more than half of the advantage of the unsmoothed low-bit attention.

\subsection{Insights from Profiling and Analysis}
\label{sec:motiv:insight}

The analysis yields three principles for low-bit attention on data-center
Blackwell GPUs.

\noindent\textbf{Principle 1: mixed precision.}
NVFP4$(QK^\top)+$FP8$(PV)$ accelerates $QK^\top$ computation without 
quantizing $P$ to NVFP4, avoiding the associated data-movement and
scale management overheads along the softmax path.

\noindent\textbf{Principle 2: reduce non-matmul overhead.}
Rank-one Tensor Core MMA compensation moves the cost from softmax to Tensor
Cores, while a fine-grained pipeline improves utilization.

\noindent\textbf{Principle 3: specialize prefill and decode.}
Prefill processes the complete query sequence and is primarily constrained by
the softmax path. Decode has only a small number of valid query rows at each
step and is primarily constrained by insufficient parallelism and padded rows
introduced by MMA atom. 

\section{Quantization Algorithm}
\label{sec:quant}

Following the first two principles in \S\ref{sec:motivation}, \SystemName uses
NVFP4 for $QK^\top$, FP8 for $PV$, and sequence centering with BF16 rank-one
Tensor Core MMA to compensate for channel-outlier-induced NVFP4 error in $Q$
and $K$.

\subsection{QK4PV8 Quantization}
\label{sec:quant:scheme}

\SystemName quantizes $Q$ and $K$ to NVFP4 with one \texttt{e4m3} scale factor
per $16$ head-dimension elements, and quantizes $P$ and $V$ to FP8. The online
softmax states $m$, $\ell$, and the $PV$ accumulator remain FP32, and the
operator returns BF16. This assignment reduces $QK^\top$ computation and KV
storage without the block-scale management required by fully NVFP4 $PV$.

Quantizing $P$ does not require a dynamically computed scale factor for each
block, but should account for the limited range of FP8 \texttt{e4m3} and the
numerical degradation caused by attention sinks. From
Equation~\ref{eq:osexp}, each unnormalized probability generated by online
softmax satisfies
\begin{equation*}
0<P^{(t)}_{ij}=e^{S^{(t)}_{ij}-m^{(t)}_i}\leq 1.
\end{equation*}
\SystemName therefore uses a fixed scale factor $s_P=448$ for $P$:
\begin{equation*}
\widehat{P}^{(t)}_{ij}
 = \operatorname{cast}_{\mathrm{e4m3}}\!\left(s_P P^{(t)}_{ij}\right) / s_P.
\end{equation*}
Because $448$ is the largest normal \texttt{e4m3} value and
$P^{(t)}_{ij}\le1$, this analytic scale applies across layers, heads, and
inputs without runtime \texttt{amax} statistics, per-block scale generation,
or calibration. The remaining concern is the lower end of the FP8 range: in a
forward traversal, an early attention sink can raise the running maximum and
push ordinary-token probabilities into the \texttt{e4m3} subnormal range.
\SystemName therefore traverses $K$ and $V$ in reverse KV-block order to delay
the sink block. Since online softmax is traversal-order invariant, reversal
changes only the finite-precision rounding path, following FA3~\cite{fa3}.

\subsection{Outlier Structure and Sequence Smoothing}
\label{sec:quant:smooth}

\begin{figure*}[t]
\centering
\begin{subfigure}[t]{0.245\textwidth}
  \centering\includegraphics[width=\linewidth]{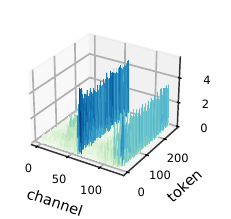}
  \caption{Llama-3.1-8B $|Q|$}\label{fig:outliers:lq}
\end{subfigure}\hfill
\begin{subfigure}[t]{0.245\textwidth}
  \centering\includegraphics[width=\linewidth]{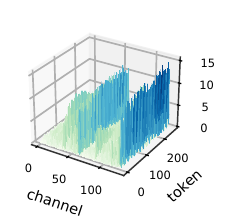}
  \caption{Llama-3.1-8B $|K|$}\label{fig:outliers:lk}
\end{subfigure}\hfill
\begin{subfigure}[t]{0.245\textwidth}
  \centering\includegraphics[width=\linewidth]{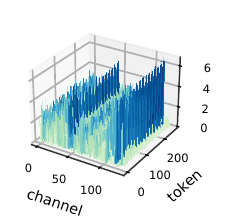}
  \caption{Qwen3-14B $|Q|$}\label{fig:outliers:qq}
\end{subfigure}\hfill
\begin{subfigure}[t]{0.245\textwidth}
  \centering\includegraphics[width=\linewidth]{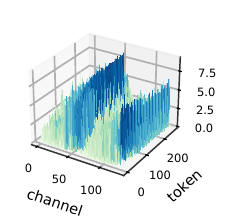}
  \caption{Qwen3-14B $|K|$}\label{fig:outliers:qk}
\end{subfigure}
\caption{Distributions of $|Q|$ and $|K|$ for one head at layer $16$ of Llama-3.1-8B and layer $20$ of Qwen3-14B.}
\label{fig:outliers}
\vspace{0mm}
\end{figure*}

NVFP4 block scaling assigns the dynamic range of each $16$-element block
according to its amax. Stable channel outliers within a block can therefore
directly reduce the effective precision of 4-bit quantization.
Figure~\ref{fig:outliers} shows activations collected from a running service:
one head from one layer of each model over the first $256$ tokens. The
magnitudes of $Q$ and $K$ form ridges along the channel dimension that persist
across many tokens, while remaining approximately flat along the token
dimension. Because NVFP4 blocks are partitioned along the channel axis, a
small number of stable outlier channels can consume most of the representable
range of their blocks. These outliers are determined by fixed model
parameters rather than transient input tokens, making them suitable for
smoothing. In contrast, $V$ varies comparably across the channel and token
dimensions and is therefore not smoothed, consistent with prior
work~\cite{sageattn2}. As $V$ is not centered, it requires no mean recovery and
is unaffected by rank-one compensation. It uses FP8 \texttt{e4m3} (three
mantissa bits versus one for NVFP4) with FP32 accumulation. Since
$O=\sum_j P_jV_j$ is a convex combination, rounding errors in $V$ enter the
output as a weighted average rather than a sum that grows linearly with
sequence length.

\noindent\textbf{Sequence centering.}
To mitigate the stable channel outlier structure, we subtract a sequence-level
mean for each layer and head before quantization. For the set of valid tokens
$\mathcal{C}$, we define
\begin{equation}
\bar{K}[h,:]=\frac{1}{|\mathcal{C}|}\sum_{t\in\mathcal{C}}K[t,h,:],
\bar{Q}[h,:]=\frac{1}{|\mathcal{C}|}\sum_{t\in\mathcal{C}}Q[t,h,:]
\label{eq:mean}
\end{equation}
The system obtains these two statistics during prefill and quantizes the
centered values
\begin{equation*}
Q_c=Q-\bar{Q},\qquad K_c=K-\bar{K}.
\end{equation*}
We partition $Q_c$, $K_c$, and $V$ along the token dimension into
$\{Q_{c,i}\}$, $\{K_{c,j}\}$, and $\{V_j\}$, with quantized blocks
$\hat{Q}_{c,i}$, $\hat{K}_{c,j}$, and $\hat{V}_j$ and scales $s_{Q,i}$ and
$s_{K,j}$. The removed means become recoverable compensation factors, while
centering is fused into the quantization write path.


\subsection{Rank-One Tensor Core Compensation}
\label{sec:quant:rank1}

When both $Q$ and $K$ are centered, the removed mean terms must be restored
before softmax. Expanding $S=QK^\top$ and defining a key-dependent
compensation vector gives
\begin{equation}
S
  = Q_cK_c^\top
   + \mathbf{1}_M\bigl(\bar{Q}K_c^\top\bigr)
   + \bigl(Q\bar{K}^\top\bigr)\mathbf{1}_N^\top
\label{eq:decomp}
\end{equation}
\begin{equation}
\Delta_S[j]
  \triangleq \bar{Q}\cdot K_c[j,:]^\top
\label{eq:delta}
\end{equation}
Here $\Delta_{S,j}$ denotes the entries of $\Delta_S[j]$ in KV block $j$;
GQA uses one $\Delta_S$ vector per query head.

The third term is constant across keys within each softmax row and cancels because
$\mathrm{softmax}(x+c\mathbf{1})=\mathrm{softmax}(x)$. Only the
key-dependent second term must be restored after centering:
\begin{equation}
\hat{S}
 = \mathrm{deq}\bigl(\hat{Q}_{c}\hat{K}_{c}^{\top}\bigr)
   + \mathbf{1}_M\Delta_S
\label{eq:shat}
\end{equation}
Here $\Delta_S$ is added before masking and before updating the running
maximum. This ordering prevents compensation from reaching masked positions
and keeps the running maximum consistent with the normalized scores.

\noindent\textbf{Rank-one MMA implementation.}
The compensation term in Equation~\ref{eq:shat} is the outer product
$\mathbf{1}_M\Delta_S$ within a score tile. Rather than adding it
element-wise after computing $\hat{Q}_c\hat{K}_c^\top$, we first issue a
BF16 Tensor Core MMA instruction on the same score accumulator:
\begin{equation}
S_{\mathrm{acc}}
  \leftarrow \mathbf{1}_M\Delta_S,
\qquad
S_{\mathrm{acc}}
  \mathrel{+}= \mathrm{deq}\bigl(\hat{Q}_{c}\hat{K}_{c}^\top\bigr).
\label{eq:seed}
\end{equation}
For the BF16 MMA reduction granularity $K{=}16$, the constant $A$ operand has
$1$ only in its first column, with the remaining columns and corresponding
$\Delta_S$ components zeroed. The accumulation is therefore exactly
$\mathbf{1}_M\Delta_S$, moving compensation from the softmax path to otherwise
idle Tensor Cores without changing online softmax. Compared with element-wise
addition, this introduces one BF16 rounding with relative error at most
$2^{-8}$.

\section{High Performance QK4PV8 FA Kernel}
\label{sec:kernel}

This section presents \SystemName's unified QK4PV8 computation flow, followed
by tensor memory reuse, rank-one compensation placement, and decode-specific
optimizations. 

\subsection{Fine-Grained Pipeline}
\label{sec:kernel:pipe}

Algorithm~\ref{alg:prefill} describes \SystemName's two-stage execution flow.
The first stage runs outside the attention operator: it computes $\bar{Q}$ and
$\bar{K}$, centers $Q$ and $K$, quantizes $Q_c$ and $K_c$ to NVFP4, quantizes
$V$ to FP8, and computes the per-token compensation vector $\Delta_S$ with
one GEMV according to Equation~\ref{eq:delta}. This work is fused with the
inference framework's KV-cache write path (\S\ref{sec:sys}), keeping its
overhead low. The second stage is the attention main loop, which consumes
only the quantized data and compensation factors.

In the algorithm, $i$ and $j$ index query blocks and KV blocks, respectively.
$\hat{Q}_{c,i}$ is the $i$-th quantized centered query block;
$\hat{K}_{c,j}$ and $\hat{V}_j$ form the $j$-th KV block; and $s_{Q,i}$ and
$s_{K,j}$ are their NVFP4 block scale factors. $S_{i,j}$ and $O_{i,j}$
denote the score and output states after processing KV block $j$, while
$m_{i,j}$ and $\ell_{i,j}$ are the running maximum and normalization factor
of online softmax. The only quantization inside the loop converts
$P_{i,j}$ to FP8. Since its range has a known upper bound
(\S\ref{sec:quant:scheme}), this conversion uses a fixed scale factor.

\begin{algorithm}[t]
\small
\caption{Algorithm design of \SystemName. $\Delta_{S,j}$ is the compensation
vector. The $1/\sqrt{d}$ factor is folded into the $\lambda$.}
\label{alg:prefill}
\begin{algorithmic}[1]
\Require $Q,K,V$ in BF16; block sizes $B_q,B_{kv}$; $T_m=\lceil N_q/B_q\rceil$
  query blocks and $T_n=\lceil N_{kv}/B_{kv}\rceil$ KV blocks; $M=B_q$
\Ensure $O$ in BF16
\State $\bar{Q}\gets\operatorname{mean}(Q),\ \bar{K}\gets\operatorname{mean}(K)$
  \Comment{outside kernel}
\State $Q_c\gets Q-\bar{Q},\ K_c\gets K-\bar{K}$
\State $(s_Q,\hat{Q}_c)\gets\operatorname{Quant}_{\mathrm{NVFP4}}(Q_c)$; \ $(s_K,\hat{K}_c)\gets\operatorname{Quant}_{\mathrm{NVFP4}}(K_c)$
\State $\hat{V}\gets\operatorname{Quant}_{\mathrm{FP8}}(V)$
  \Comment{fused into KV-cache write}
\State $\Delta_S\gets\textsc{GEMV}_\mathrm{bf16}\bigl(\bar{Q},K_c^\top\bigr)$
  \Comment{Equation~\ref{eq:delta}}
\For{$i=1$ to $T_m$}
  \Comment{kernel main loop}
  \State $m_{i,0}\gets-\infty,\ \ell_{i,0}\gets0,\ O_{i,0}\gets0$
  \For{$j=1$ to $T_n$}
    \State load $\hat{Q}_{c,i},s_{Q,i},\hat{K}_{c,j},s_{K,j},
      \hat{V}_j,\Delta_{S,j}$
    \Comment{Equation~\ref{eq:seed}}
    \State {\mbox{$S_{i,j} \gets
      \textsc{MMA}_{\mathrm{bf16}}(\mathbf{1}_M,\Delta_{S,j})+
      \textsc{MMA}_{\mathrm{fp4}}(\hat{Q}_{c,i},s_{Q,i},
      \hat{K}_{c,j},s_{K,j})$}}
    \State $m_{i,j}\gets\max\bigl(m_{i,j-1},\operatorname{rowmax}(S_{i,j})\bigr)$
    \State $P_{i,j}\gets 2^{\lambda(S_{i,j}-m_{i,j})}$,\ \
      $\lambda=\log_2\!e/\sqrt{d}$
      \Comment{$1/\sqrt{d}$ folded here}
    \State $\ell_{i,j}\gets
      2^{\lambda(m_{i,j-1}-m_{i,j})}\ell_{i,j-1}
      +\operatorname{rowsum}(P_{i,j})$
      \Comment{FP32 $P_{i,j}$}
    \State $\hat{P}_{i,j}\gets
      \operatorname{Quant}_{\mathrm{FP8}}(P_{i,j})$
    \State $O_{i,j}\gets
      \operatorname{diag}\!\left(2^{\lambda(m_{i,j-1}-m_{i,j})}\right)O_{i,j-1}
      +\textsc{MMA}_{\mathrm{fp8}}(\hat{P}_{i,j},\hat{V}_j)$
  \EndFor
  \State $O_i\gets\operatorname{diag}(\ell_{i,T_n})^{-1}O_{i,T_n}$
\EndFor
\State \Return $O=\{O_i\}_{i=1}^{T_m}$
\end{algorithmic}
\end{algorithm}

\begin{figure*}[t]
\centering
\includegraphics[width=\textwidth]{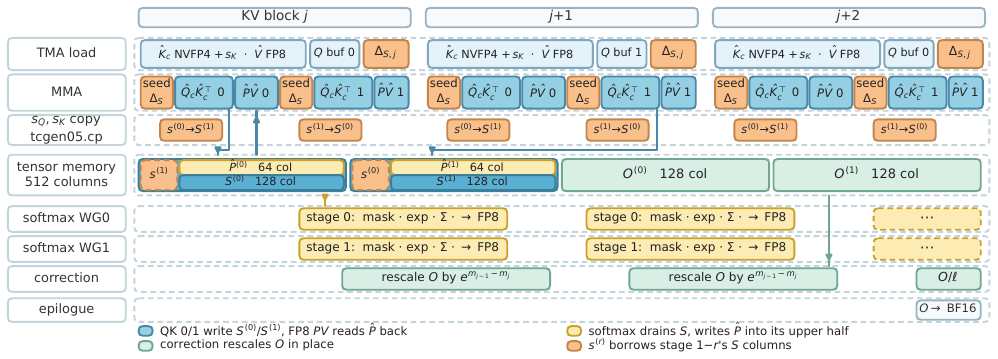}
\caption{\SystemName's steady-state pipeline. The seven execution channels for five
warp roles within one CTA advance over consecutive KV blocks, together with
their tensor memory accesses.}
\label{fig:pipeline}
\vspace{1mm}
\end{figure*}

\noindent\textbf{Warp specialization.}
\SystemName retains the five warp roles of FA4~\cite{fa4}, but adapts them for
QK4PV8: load, MMA, softmax, correction, and epilogue,
as shown in Figure~\ref{fig:pipeline}. The load warp issues asynchronous
copies of $\hat{K}_{c,j}$, $s_{K,j}$, $\hat{V}_j$, the current query
block and its scale factor, and $\Delta_{S,j}$ from global memory to on-chip
memory. The MMA warp issues two MMAs for the current score
state: a rank-one BF16 MMA first writes $\Delta_{S,j}$, followed by a
block-scaled NVFP4 MMA that accumulates
$\hat{Q}_{c,i}\hat{K}_{c,j}^{\top}$; an FP8 MMA then computes
$\hat{P}_{i,j}\hat{V}_j$. The two softmax warpgroups apply the mask, update
$m_{i,j}$ and $\ell_{i,j}$, and convert $P_{i,j}$ to $\hat{P}_{i,j}$ in
place. The correction warpgroup rescales the output state and performs final
normalization, while the epilogue warp writes the BF16 output
back to global memory.

We double-buffer the query blocks (${Q}_{c}$) to overlap MMA with softmax. 
One CTA processes
two query stages, indexed by $r\in\{0,1\}$. Each stage owns one $S^{(r)}$
and one $O^{(r)}$, corresponding to an $(S_{i,j},O_{i,j})$ state in the
algorithm. The probability matrix $\hat{P}_{i,j}$ receives no separate
tensor-memory allocation; it is written into the upper half of the current
$S^{(r)}$ region, and the FP8 $PV$ MMA reads it back from that region.
Thus, stage $r$'s NVFP4 $QK^\top$ writes only $S^{(r)}$, its softmax reads
$S^{(r)}$ and writes $\hat{P}^{(r)}$ in the upper half, and its FP8 $PV$
MMA reads $\hat{P}^{(r)}$ from the same region. The two stages alternate
between $S^{(0)}/\hat{P}^{(0)}$ and $S^{(1)}/\hat{P}^{(1)}$, allowing one
stage's softmax to overlap with the other stage's matrix multiplication.
This double buffer hides softmax latency and enables reuse of the scale factors
described next.

\noindent\textbf{Asynchronous pipeline for scale factors in tensor memory.}
Each CTA is allocated $512$ tensor memory columns. The score and output
states of the two query stages require $128$ columns each, so the four
accumulators consume the entire capacity. NVFP4 scale factors for $Q$ and
$K$ cannot receive independent allocations. \SystemName resolves this conflict by
exploiting asymmetric lifetimes: $S^{(r)}$ and $O^{(r)}$ must remain resident
through a long softmax or main-loop interval, whereas $s_{Q,i}$ and $s_{K,j}$
are read only when the corresponding NVFP4 MMA is issued.

We asynchronously place FP8 scale factors in reused storage. The scale
factors needed by stage $r$ are not placed on its own $S^{(r)}$ region;
instead, they borrow columns from $S^{(1-r)}$: $s^{(0)}$ occupies the
starting columns of $S^{(1)}$, and $s^{(1)}$ occupies those of $S^{(0)}$
(the dashed cells at the left of the score regions in Figure~\ref{fig:pipeline}).
The double-buffer timing makes this safe. While stage $r$ issues $QK^\top$,
stage $1-r$'s score region is in the gap after softmax and $PV$ have
consumed it but before the next MMA overwrites it. In contrast, $S^{(r)}$
is the target of the current MMA and cannot simultaneously hold its scale
factors.

Scale-factor reads and writes form an independent pipeline with a phase
distinct from data loading and matrix multiplication. Each query
stage owns a set of \texttt{tcgen05.cp} copies and its own barrier phase
(the two $s_Q,s_K$ copy boxes in Figure~\ref{fig:pipeline}). Copies are
issued early enough for the factors to arrive before the corresponding MMA;
their transfer overlaps with the other stage's softmax and matrix
multiplication and therefore does not enter the critical path. Correctness
follows from two phase conditions: scale factors must arrive before their
MMA is issued, and a new write to a score region can occur only after that
region has been consumed by softmax and the $PV$ MMA. The probability matrix
$P$ uses the same lifetime reuse but occupies only the upper half of the
score region, requiring no additional tensor-memory allocation. This allows
scale factors, attention scores, and outputs to remain resident within the
fixed $512$-column budget.

\noindent\textbf{Other attention optimizations.}
\SystemName builds the kernel on the CUTLASS and CuTe programming
primitives. 
For causal attention, \SystemName balances the triangular workload using
longest-processing-time-first scheduling and overlaps TMA and MMA with a
ping-pong KV pipeline.
It also distributes softmax exponentiation between a
polynomial approximation and hardware $2^x$ instructions to balance the FMA
and SFU pipelines.
Moreover, attention scaling is fused into exponentiation:
$Q$ and $K$ remain in the unscaled inner-product domain, and the softmax
constant $\lambda=\log_2\!e/\sqrt{d}$ is folded into the exponent
($2^{\lambda(S-m)}=2^{\lambda S-\lambda m}$), avoiding separate scaling of
the scores and compensation term $\Delta_{S,j}$.

\subsection{MMA Compensation}
\label{sec:kernel:rank1}

For each $(i,j)$, the MMA warp first issues
\textsc{MMA}$_{\mathrm{bf16}}(\mathbf{1}_M,\Delta_{S,j})$, then accumulates
the NVFP4
\textsc{MMA}$_{\mathrm{nvfp4}}(\hat{Q}_{c,i},s_{Q,i},
\hat{K}_{c,j},s_{K,j})$ into the same score state $S_{i,j}$. Because the
compensation has already been written to the score accumulator before
softmax, the softmax warp reads complete scores without an additional
element-wise addition or a change to the online-softmax instruction sequence.

The compensation vector $\Delta_{S,j}$ is much smaller than
$\hat{K}_{c,j}$ and $\hat{V}_j$. The loading warp prefetches it and publishes
it with the same barrier phase as the current KV block. Rank-one MMA and
NVFP4 $QK^\top$ consequently share one dependency and require no additional
synchronization. Relative to the same build without smoothing, the prefill
attention operator adds only $1.3$--$2.5\%$; placing compensation on the
softmax warp costs $18.6$--$22.7\%$ (\S\ref{sec:eval:ablation}).


\subsection{Decode-Specific Optimizations}
\label{sec:kernel:decode}

Decode uses the same representations and MMA precisions as prefill but is
parallelism-bound. \SystemName packs
GQA heads, partitions the KV dimension, removes softmax work from padded rows,
and merges the resulting local states.

GQA lets $H_q/H_{kv}$ query heads share one KV head, so \SystemName packs
each group into the MMA $M$ dimension and assigns one (request, KV-head) pair
to each CTA. It partitions the history into fixed $8\times128$-token chunks;
each CTA maintains local $(m,\ell,O)$ states in preallocated workspaces,
making the path CUDA Graph capturable and increasing parallelism over
split-KV-disabled BF16 FA4.

For GQA $32{:}8$, FP4 MMA fixes $M=128$ although each group has only four real
rows. \SystemName replicates these rows and processes their corresponding
columns as four diagonal $32\times32$ tiles, reducing softmax work from
$32\times128$ to $32\times32$ per row range and the full-tile $2^x$
evaluations from $128\times128$ to $128\times32$. The four warps produce four
states per partition, yielding $4N_p$ states merged by log-sum-exp. Eight
independent state streams and shared-memory scale prefetching hide the
reduction's memory latency; the same strategy extends to other model shapes.

\section{LLM Serving System}
\label{sec:sys}

\noindent\textbf{Serving system.}
To evaluate end-to-end performance, we integrate \SystemName into
SGLang~\cite{sglang} as an attention backend, following prior
work~\cite{fasttree}. The backend supports paged prefill and decode, GQA,
masking, and CUDA Graph capture, while SGLang handles the remaining model
execution. During decode, fixed-size cache pages and preallocated tensors keep
the number of KV partitions constant after graph capture. \SystemName returns
BF16 outputs through the standard attention interface. With smoothing, the
per-token compensation $\Delta_S$ follows the KV cache's physical page-table
indices and resides in a side cache directly accessed during decode.

\noindent\textbf{Online quantization.}
\SystemName fuses NVFP4 quantization of $Q$, low-bit $K/V$ cache writes, and
compensation generation into SGLang's cache-write path. It stores $K$ in
NVFP4 with one \texttt{e4m3} scale byte per $16$ elements and $V$ in FP8,
allowing direct consumption without dequantization. For equal-dimensional
$K$ and $V$, the layout uses $1/2.56$ of the BF16 KV-cache footprint:
$1/2+1/16$ bytes per $K$ element and one byte per $V$ element, versus two
bytes per element for each BF16 tensor.
With smoothing, prefill computes per-layer, per-head $\bar{Q}$ and $\bar{K}$
over valid tokens and freezes them for the sequence lifetime. Decode centers
only the newly appended token against these per-request means without
recomputing them. This avoids degenerate single-token query centering that
would shift the entire score correction to $\Delta_S$. For the new token,
centering and compensation generation are fused into the cache-write kernel. 
Without smoothing, no side cache is allocated. Kernel measurements exclude
quantization statistics and compensation generation; end-to-end measurements
include quantization, sequence centering, and KV-cache writes.

\section{Evaluation}
\label{sec:eval}

\begin{figure*}[t]
\centering
\includegraphics[scale=1]{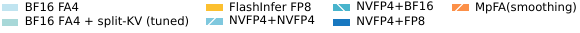}\\[-1pt]
\begin{subfigure}[t]{0.56\textwidth}
  \centering\includegraphics[width=\linewidth]{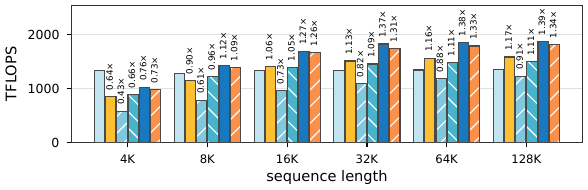}
  \caption{causal prefill}\label{fig:kperf:pre}
\end{subfigure}\hfill
\begin{subfigure}[t]{0.42\textwidth}
  \centering\includegraphics[width=\linewidth]{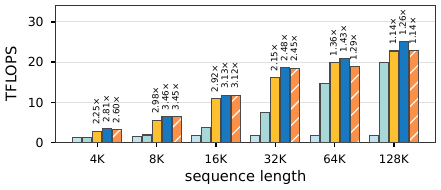}
  \caption{decode}\label{fig:kperf:dec}
\end{subfigure}
\caption{Attention-kernel throughput on an SM100 GPU with GQA $32{:}8$ and
$head\_dim{=}128$, the shape of Llama-3.1-8B. Speedups above
the bars use BF16 FA4 in (a) and \emph{split-KV-enabled} BF16 FA4 in (b).}
\label{fig:kperf}
\end{figure*}

\begin{figure*}[t]
\centering
\includegraphics[scale=1]{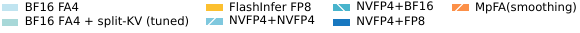}\\[-1pt]
\begin{subfigure}[t]{0.56\textwidth}
  \centering\includegraphics[width=\linewidth]{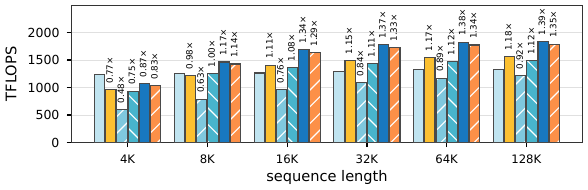}
  \caption{causal prefill}\label{fig:kperf14b:pre}
\end{subfigure}\hfill
\begin{subfigure}[t]{0.42\textwidth}
  \centering\includegraphics[width=\linewidth]{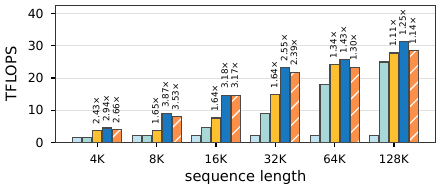}
  \caption{decode}\label{fig:kperf14b:dec}
\end{subfigure}
\caption{Attention-kernel throughput for Qwen3-14B (GQA $40{:}8$). The
baselines and speedup denominators match Figure~\ref{fig:kperf}.}
\label{fig:kperf14b}
\end{figure*}

\subsection{Experimental Setup}
\label{sec:eval:setup}

\noindent\textbf{Hardware and software.}
We conduct all experiments on a single NVIDIA SM100 GPU using CUDA $13.0$,
PyTorch $2.11.0$, and SGLang $v0.5.15$. For end-to-end experiments, we enable CUDA
Graph capture and disable Radix Cache~\cite{sglang}.


\noindent\textbf{Models and accuracy tasks.}
We evaluate Llama-3.1-8B and Qwen3-14B, covering GQA $32{:}8$ and $40{:}8$;
both have 128-dimensional heads. We extend Qwen3-14B from its native $40$K to $128$K using YaRN, while
Llama-3.1-8B uses its native $128$K context. This extension may affect
accuracy but not performance measurements.

We evaluate MMLU~\cite{mmlu} (5-shot, all $57$ subjects), HellaSwag~\cite{HellaSwag}
(20-shot), WinoGrande~\cite{WinoGrande} (0-shot), GSM8K~\cite{gms8k} (5-shot with
generated reasoning), and RULER-QA~\cite{ruler}, which pads SQuAD and HotpotQA
passages to $64$K tokens. Each backend uses the same examples and prompts with
greedy decoding ($temperature=0$).

\noindent\textbf{Baselines.}
We compare BF16 FA4, FP8 FlashInfer, and three precision assignments for
$QK^\top$ and $PV$: NVFP4$+$NVFP4, NVFP4$+$BF16, and NVFP4$+$FP8.
\SystemName is the complete QK4V8 design with rank-one smoothing. For decode
kernel tests, we tune BF16 FA4's split count per context length and use
FlashInfer's default split-KV policy. \SystemName instead uses fixed
KV-partition and column-split configurations. For end-to-end tests, BF16 FA4
runs one CTA per request because its split-KV path preallocates workspaces per
\texttt{num\_splits} and cannot be captured by CUDA Graph. FlashInfer and
\SystemName use graph-compatible split-KV, with FlashInfer using an FP8 KV
cache. Integrated as an SGLang attention backend, \SystemName supports the same
production features as the baselines: continuous batching, paged KV cache,
CUDA Graph capture, and \texttt{head\_dim} ($128$ for both models).

We measure prefill/decode kernel throughput, time to first token (TTFT), time
per output token (TPOT), output-token throughput (OTPS) across concurrency
levels, end-to-end accuracy, and ablations. Following prior
work~\cite{armsme}, we report the mean of five performance runs.

\begin{figure*}[t]
\centering
\includegraphics[scale=1]{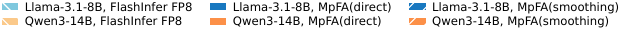}\\[-1pt]
\begin{subfigure}[t]{0.49\textwidth}
  \centering\includegraphics[width=\linewidth]{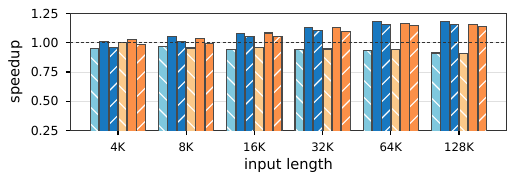}
  \caption{Time to first token (TTFT)}\label{fig:e2elat:ttft}
\end{subfigure}\hfill
\begin{subfigure}[t]{0.49\textwidth}
  \centering\includegraphics[width=\linewidth]{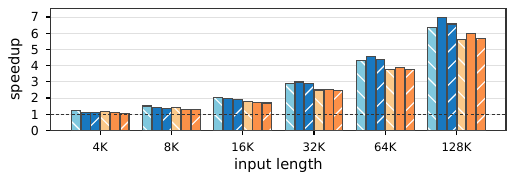}
  \caption{Time per output token (TPOT)}\label{fig:e2elat:tpot}
\end{subfigure}
\caption{End-to-end latency speedup over BF16 FA4 at concurrency $1$.}
\label{fig:e2elat}
\end{figure*}

\subsection{FlashAttention Kernel Performance}
\label{sec:eval:kernel}

Figure~\ref{fig:kperf} compares causal prefill and decode throughput across
precision configurations, using BF16 FA4 and split-KV-enabled BF16 FA4 as
their respective denominators. NVFP4 $+$ FP8 is the hardware-efficient
prefill assignment; decode also depends on KV partitioning and softmax
scheduling. \SystemName includes smoothing, while direct NVFP4 $+$ FP8
isolates its compensation cost.

\noindent\textbf{Prefill workloads.}
At $16$K and above, measured throughput follows direct NVFP4 $+$ FP8 $>$ \SystemName $>$
NVFP4 $+$ BF16 $>$ BF16 $>$ NVFP4 $+$ NVFP4, consistent with
\S\ref{sec:motiv:gap}. Across $16$K--$128$K, \SystemName sustains
$1672$--$1817$\,TFLOPS and achieves a $1.31\times$ geometric-mean speedup
over BF16 FA4; the direct variant reaches $1.35\times$, showing that smoothing
preserves most of the gain. FP8 FlashInfer reaches $1.12\times$ over BF16,
while \SystemName is $1.17\times$ faster than FlashInfer. The benefit follows
from reducing the compute-intensive $QK^\top$ from FP8 to NVFP4. Qwen3-14B's
GQA $40{:}8$ shape shows the same trend (Figure~\ref{fig:kperf14b}), with
$1.33\times$/$1.15\times$ speedups over BF16 FA4/FP8 FlashInfer.

Fixed overhead limits short sequences: at $4$K, \SystemName and its direct
variant reach only $0.74\times$ and $0.76\times$ of BF16, versus
$1.09\times$ and $1.12\times$ at $8$K. The crossover therefore falls between
$4$K and $8$K, below which matrix multiplication does not dominate.
\SystemName is thus suited to end-to-end inference at sequence lengths
beyond $8$K.
NVFP4 $+$ BF16 reaches $1.05$--$1.11\times$ at $\ge16$K, confirming the
benefit of quantizing $QK^\top$ alone. Relative to FP8, however, its $PV$
MMA and shared-memory traffic for $V$ both double, and it cannot use a
low-bit $V$ cache. Fully NVFP4 is slowest from $4$K onward, reaching only
$0.87$--$0.90\times$ BF16 at long sequences because of the additional
$PV$ execution-path costs identified in \S\ref{sec:motiv:gap}.

\noindent\textbf{Decode workloads.}
At $128$K, \SystemName and its direct variant reach only $22.8$ and
$25.1$\,TFLOPS, respectively, confirming that long-context decode is
parallelism-bound rather than Tensor-Core-bound. We compare against three
baselines with distinct deployment constraints.

The first is production-deployable BF16 FA4. Its split-KV path preallocates
workspaces according to \texttt{num\_splits} and cannot be captured by CUDA
Graph, so SGLang uses one CTA to process the full history sequentially.
Relative to this configuration, \SystemName improves throughput from
$2.63\times$ at $4$K to $13.04\times$ at $128$K; the direct variant reaches
$2.83\times$ and $14.37\times$, respectively.

The second is BF16 FA4 with split-KV enabled. We sweep
\texttt{num\_splits}$\in\{1,\dots,32\}$ and select the best value at each
length; the optimum is $16$ or $32$ at $\ge8$K. Against this per-length-tuned
baseline, \SystemName achieves a $2.08\times$ average speedup
across $8$K--$128$K without per-length tuning, compared with $2.17\times$ for
the direct variant. \SystemName peaks at $3.45\times$ at $8$K and remains
$1.14\times$ faster at $128$K, due to GQA packing, fixed KV partitions,
column splitting, and parallel state merging. The speedup decreases with
sequence length: partitioning exposes substantially more parallelism than the
baseline at short-to-medium lengths, whereas longer KV histories make both
systems HBM-bandwidth-bound, limiting the benefit of NVFP4 compute.

The third is FlashInfer's FP8 Blackwell FlashAttention, which supports split-KV
under CUDA Graph. \SystemName achieves a $1.06\times$ geometric-mean speedup
across $8$K--$128$K, leading by $1.11$--$1.14\times$ through $32$K but
falling to $0.95\times$ at $64$K and near parity ($0.99\times$) at $128$K.
The direct variant reaches $1.11\times$ in geometric mean and remains
$1.05\times$/$1.09\times$ faster at $64$K/$128$K. The gap reflects smoothing:
the $\Delta_S$ side cache adds about $4\%$ KV traffic, which is mostly absorbed
by L2 through $32$K but reaches HBM beyond $64$K, making \SystemName about
$10\%$ slower than the direct variant.
The modest margin over FlashInfer also reflects the decode regime: at
concurrency $1$, reading the KV history dominates, so NVFP4's compute
advantage over FP8 is largely unused. Moreover, block-scaled FP4
MMA requires $M{=}128$, while a GQA group has only $H_q/H_{kv}$ real query
rows; row replication and column splitting cannot eliminate all padded work.
These decode results generalize to GQA $40{:}8$ (Figure~\ref{fig:kperf14b}):
\SystemName peaks at $3.53\times$ over split-KV BF16 FA4 at $8$K and stays
$1.14\times$ at $128$K. Its larger query-to-KV ratio feeds more compute per KV
read, yielding $0.97\times$/$1.02\times$ of FP8 FlashInfer at $64$K/$128$K.

\begin{table*}[t]
\centering
\small
\setlength{\tabcolsep}{6pt}
\renewcommand{\arraystretch}{1.25}
\caption{End-to-end accuracy. Rec.\ is the fraction of NVFP4$+$FP8 loss
(vs.\ BF16) that \SystemName recovers.}
\label{tab:acc}
\begin{tabular}{l|ccccr|ccccr}
\hline
 & \multicolumn{5}{c|}{Llama-3.1-8B} & \multicolumn{5}{c}{Qwen3-14B} \\
Dataset & BF16 & FP8 & NVFP4$+$FP8 & \SystemName & Rec.
        & BF16 & FP8 & NVFP4$+$FP8 & \SystemName & Rec. \\
\hline
MMLU       & 0.685 & 0.683 & 0.644 ($-$0.041) & \textbf{0.674} ($-$0.011) & 73\%
           & 0.787 & 0.787 & 0.776 ($-$0.011) & \textbf{0.785} ($-$0.002) & 82\% \\
HellaSwag  & 0.785 & 0.785 & 0.765 ($-$0.020) & \textbf{0.780} ($-$0.005) & 75\%
           & 0.774 & 0.774 & 0.768 ($-$0.006) & \textbf{0.773} ($-$0.001) & 83\% \\
WinoGrande & 0.752 & 0.751 & 0.735 ($-$0.017) & \textbf{0.745} ($-$0.007) & 59\%
           & 0.727 & 0.726 & 0.712 ($-$0.015) & \textbf{0.720} ($-$0.007) & 53\% \\
GSM8K      & 0.761 & 0.757 & 0.653 ($-$0.108) & \textbf{0.736} ($-$0.025) & 77\%
           & 0.887 & 0.882 & 0.877 ($-$0.010) & \textbf{0.882} ($-$0.005) & 50\% \\
RULER-QA   & 0.624 & 0.619 & 0.533 ($-$0.091) & \textbf{0.578} ($-$0.047) & 49\%
           & 0.656 & 0.656 & 0.592 ($-$0.064) & \textbf{0.607} ($-$0.049) & 24\% \\
\hline
\end{tabular}
\end{table*}

\begin{figure}[t]
\centering
\includegraphics[scale=1]{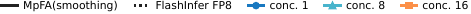}\\[-1pt]
\includegraphics{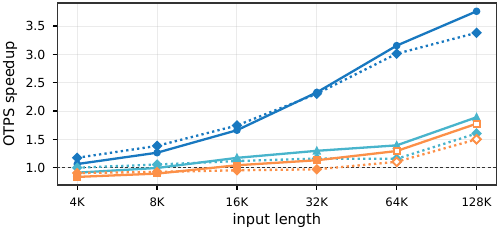}
\caption{Output throughput of SGLang serving relative to BF16 FA4 for
Qwen3-14B.}
\label{fig:e2ethr}
\vspace{-4mm}
\end{figure}

\subsection{End-to-End Inference Performance}
\label{sec:eval:e2e}

\noindent\textbf{Latency.}
Figure~\ref{fig:e2elat} compares \SystemName, its direct variant, BF16 FA4,
and FP8 FlashInfer under identical serving parameters. Across both models and
$16$K--$128$K, \SystemName improves TTFT and TPOT over BF16 by $1.11\times$
and $3.30\times$ in geometric mean. At $128$K, TPOT falls from $42.4$ to
$6.5$\,ms for Llama-3.1-8B ($6.58\times$) and from $53.8$ to $9.5$\,ms for
Qwen3-14B ($5.67\times$), slightly exceeding FlashInfer's
$6.38\times$/$5.59\times$.

Prefill gains are smaller because non-attention computation remains
unchanged: at $128$K, \SystemName improves TTFT by $1.16\times$/$1.14\times$.
Although FP8 FlashInfer outperforms BF16 FA4 (Figure~\ref{fig:kperf}), separately producing its FP8 KV cache offsets this
kernel gain end to end, giving $0.91$--$0.96\times$ BF16 TTFT. \SystemName
fuses sequence centering, quantization, and cache writes
(Figure~\ref{fig:pass}), improving TTFT by up to $1.16\times$ over BF16 FA4 and
$1.16$--$1.27\times$ over FP8 FlashInfer at $32$K--$128$K.

Fixed overhead limits both latency metrics at short sequences, so we make no
stable speedup claim for inputs of $\le2$K. Relative to \SystemName(direct),
smoothing adds $1.5$--$7.3\%$ to TTFT and $3.0$--$6.2\%$ to TPOT over
$4$K--$128$K; reduced KV traffic dominates as context grows.

\noindent\textbf{Throughput.}
Figure~\ref{fig:e2ethr} reports output throughput relative to end-to-end
BF16 FA4 with fixed input length and $1024$ output tokens. At concurrency $1$,
\SystemName achieves a $2.81\times$ geometric-mean speedup across both models
and $16$K--$128$K, reaching $4.47\times$/$3.76\times$ at $128$K
(Llama-3.1-8B/Qwen3-14B). Absolute throughput then increases from
$21.2$ to $94.8$\,tok/s and from $16.2$ to $61.0$\,tok/s. The smaller
throughput gain than TPOT reflects the more modest $1.14$--$1.16\times$ TTFT
speedup. This speedup is measured against the production-deployable BF16 FA4,
whose split-KV path cannot be graph-captured. However, the controlled kernel
comparison with per-length split-KV tuning (Figure~\ref{fig:kperf:dec}) shows
that \SystemName remains faster at every length ($2.08\times$ on average and
$1.14\times$ at $128$K), confirming that split-KV does not change the
end-to-end performance ordering.

Relative to FP8 FlashInfer,
\SystemName leads across $32$K--$128$K by a $1.06\times$ geometric mean,
reaching $1.11$--$1.12\times$ at $128$K on both models. The margin grows with context
as sequence centering, quantization, and KV cache writes amortize and
\SystemName's lower KV cache traffic comes to dominate.

Higher concurrency narrows \SystemName's advantage over BF16 but preserves the
long-context benefit: at concurrency $8$, it achieves a $1.33\times$
average speedup across both models and $8$K--$128$K, reaching
$1.70\times$/$1.89\times$ at $128$K. For Qwen3-14B at concurrency $16$, it
overtakes BF16 at $16$K ($1.04\times$) and reaches $1.13\times$ at the fully
resident $32$K point. Batching supplies BF16 with parallelism, while the $2.56\times$ KV-byte
reduction persists.

The trend relative to FlashInfer holds under batching: at concurrency $8$ the
advantage is a $1.12\times$ geometric mean across $8$K--$128$K, from parity at
$8$K to $1.18$--$1.31\times$ at $128$K; for Qwen3-14B at concurrency $16$ it is
$1.09\times$/$1.17\times$ at the fully resident $16$K/$32$K points. Points
beyond $32$K at concurrency $16$ are capacity constrained because the KV pool
admits only $11$ and $5$ requests at $64$K and $128$K, so we exclude them from
aggregate results.

\noindent\textbf{Accuracy.}
Table~\ref{tab:acc} compares four backends across five benchmarks and two
models. FP8 FlashInfer stays within $0.006$ of BF16 for every model--task pair.
Direct NVFP4 $+$ FP8 loses most on Llama-3.1-8B GSM8K ($0.108$) and RULER-QA
($0.091$); \SystemName recovers $62.5\%$ of the loss on average
($24\%$--$83\%$ per pair). It recovers $49\%$--$77\%$ on Llama-3.1-8B,
trailing FP8 by at most $0.009$ on knowledge and commonsense tasks. Qwen3-14B
has smaller direct losses ($0.006$--$0.015$ on single-token tasks), with
$50\%$--$83\%$ recovery on the four shorter tasks and $24\%$ on RULER-QA.
Across both models the residual concentrates on RULER-QA and GSM8K, where
RULER-QA recovery is $24\%$--$49\%$: at $64$K, error accumulated across many
attention steps is less amenable to per-channel rank-one compensation.
\SystemName thus trades a few points on long-context and
multi-step reasoning for a $2\times$ lower-precision $QK^\top$ path that is
faster than FP8 FlashInfer (\S\ref{sec:eval:kernel}), while restoring most of
the accuracy loss without retraining. 

\subsection{Ablation Study}
\label{sec:eval:ablation}

\begin{figure}[t]
\centering
\includegraphics[scale=1]{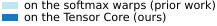}\\[-1pt]
\begin{subfigure}[t]{0.49\columnwidth}
  \centering\includegraphics[width=\linewidth]{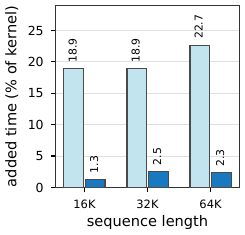}
  \caption{Relative overhead}\label{fig:ablation:rel}
\end{subfigure}\hfill
\begin{subfigure}[t]{0.49\columnwidth}
  \centering\includegraphics[width=\linewidth]{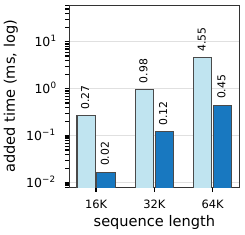}
  \caption{Absolute overhead}\label{fig:ablation:abs}
\end{subfigure}
\caption{Cost of the two compensation placements for causal prefill with
the Llama-3.1-8B shape.}
\label{fig:ablation}
\vspace{-4mm}
\end{figure}

\begin{figure}[t]
\centering
\includegraphics{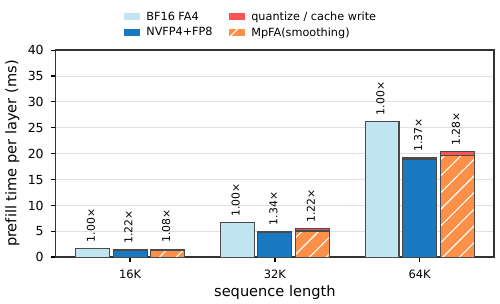}
\caption{Per-layer prefill cost, including attention and the quantization /
cache-write pass (Llama-3.1-8B).}
\label{fig:pass}
\vspace{-4mm}
\end{figure}

\noindent\textbf{Compensation placement.}
Figure~\ref{fig:ablation} compares two placements of the same compensation
term $\Delta_S$ in QK4PV8 FA4. Staging compensation in shared memory,
synchronizing after each KV block, and adding it element-wise incurs $20.1\%$
geometric-mean overhead across $16$K--$64$K. With rank-one compensation, BF16
MMA first writes to tensor memory and NVFP4 $QK^\top$ accumulates into it,
reducing overhead to $2.0\%$. At $64$K, absolute overhead falls from
$4.55$\,ms to $0.45$\,ms, or from more than half of the direct kernel's
prefill advantage to less than one tenth.
Both placements compute the same term: element-wise placement uses FP32 scalar
addition, whereas rank-one placement carries $\Delta_S$ in BF16 and accumulates
into FP32, adding one BF16 rounding with relative error no larger than $2^{-8}$.
Table~\ref{tab:acc} includes this effect because \SystemName uses rank-one
compensation by default.

\noindent\textbf{Quantization overhead.}
Quantization, sequence centering, and compensation generation remain part of
end-to-end inference; \SystemName fuses them with cache writes. Figure~\ref{fig:pass}
includes this pass in the per-layer prefill cost and labels the resulting
speedup over BF16 FA4. Across $32$K and $64$K, \SystemName achieves a
$1.25\times$ geometric-mean speedup, compared with $1.34\times$ for direct
NVFP4 $+$ FP8, preserving most of the kernel gain.

\section{Related Work}
\label{sec:related}

FlashAttention~\cite{fa1,fa2} reduces global-memory traffic through tiling
and online softmax, and later work optimizes parallelism and data reuse for
different GPUs. FlashAttention-3 (FA3)~\cite{fa3} adds warp specialization on
SM90 (Hopper) GPUs to overlap matrix multiplication and softmax across warps, and
FlashAttention-4 (FA4)~\cite{fa4} redesigns the pipeline for Blackwell tensor
memory, fifth-generation Tensor Cores, and asymmetric hardware scaling, serving
as our direct SM100 baseline. Other work adapts FlashAttention to the cache and
data-reuse characteristics of Arm multicore CPUs~\cite{armfa}, while a parallel
line designs dedicated attention accelerators, such as the approximation- and
pruning-based sparse-attention architectures SpAtten~\cite{spatten} and
ELSA~\cite{elsa}. Collectively,
these systems show that attention performance depends on
co-designing algorithmic tiling with the hardware data path rather than
optimizing solely for matrix multiplication peak throughput. \SystemName
retains FlashAttention's online softmax and FA4's SM100 execution model but
addresses a different problem. While FA4 primarily uses BF16, \SystemName
moves $QK^\top$ and $PV$ to low-bit matrix multiplication paths and analyzes
how block scaling, tensor memory capacity, and operand delivery jointly
constrain the precision assignment.

Prior low-bit LLM quantization methods, including
MR-GPTQ~\cite{fp4ht}, AWQ~\cite{awq}, GPTQ~\cite{gptq}, and
SmoothQuant~\cite{smoothquant}, target weights and activations in
linear layers rather than attention activations.
SageAttention~\cite{sageattn} studies post-training
quantization for attention: it smooths channel outliers in $K$, quantizes
$QK^\top$ to INT8, and keeps the more error-sensitive $PV$ in FP16 with an
FP16 accumulator. SageAttention2~\cite{sageattn2} further quantizes
$QK^\top$ to INT4, represents $P$ and $V$ in FP8, and applies more extensive
outlier smoothing to recover accuracy. These systems share a post-training,
plug-and-play approach and do not require both matrix multiplications to use
the same precision.
SageAttention3~\cite{sageattn3} extends this approach
to consumer Blackwell GPUs, using NVFP4 matrix multiplication
for both $QK^\top$ and $PV$, together with $Q$/$K$ smoothing and probability
quantization.
HiFA4~\cite{hifa4} also provides training-free 4-bit attention,
but targets the Ascend HIF4 NPU and its platform-specific formats and
execution units. These systems do not address tensor memory, operand delivery,
or asymmetric scaling between matmul and non-matmul units on data-center
Blackwell GPUs. KV-cache quantization methods such as KIVI~\cite{kivi} lower
cache precision to save memory and traffic, complementing quantized attention
that consumes a low-bit cache without dequantization. At the serving level,
schedulers exploiting the distinct prefill and decode profiles---through phase
disaggregation~\cite{splitwise} or prefill-decode overlap~\cite{podattn}---are
orthogonal to \SystemName, which accelerates the attention operator each phase
relies on.

Different from prior work, \SystemName is a high performance QK4PV8 attention kernel for
data-center Blackwell GPUs. We characterize the SM100 data path and hardware
behavior of low-bit attention, build mixed-precision pipelines specialized
for prefill and decode, and integrate them into an end-to-end serving system.
\SystemName further applies rank-one Tensor Core MMA smoothing compensation,
which recovers accuracy without placing additional compensation work on the
already saturated softmax path.

\section{Conclusion}
\label{sec:conclusion}

We present \SystemName, a training-free QK4PV8 FlashAttention kernel optimized for 
data-center Blackwell GPUs. 
Despite SM100's substantially higher 4-bit Tensor Core throughput, 
the overheads of online quantization, data movement, and tensor memory contention make 
fully NVFP4 attention inefficient. 
\SystemName instead adopts mixed-precision NVFP4$(QK^\top)$ + FP8$(PV)$ and addresses 
its performance challenges 
through a fine-grained asynchronous pipeline, tensor memory reuse, and 
stage-specific optimizations for prefill and decode.
To reduce quantization error without further stressing the softmax path, 
\SystemName introduces rank-one Tensor Core MMA smoothing compensation. 
On SM100, this compensation recovers $62.5\%$
of the accuracy loss with only $2.0\%$ average kernel overhead. Across
$16$K--$128$K contexts, \SystemName achieves average prefill speedups of
$1.31$--$1.33\times$ over BF16 FA4 and $1.15$--$1.17\times$ over FP8
FlashInfer for GQA $32{:}8$ and $40{:}8$, and achieves $2.81\times$ the
end-to-end output throughput of BF16 FA4 across both evaluated models.

\bibliographystyle{ACM-Reference-Format}
\bibliography{refs}

@inproceedings{fa1,
  author    = "Tri Dao and Daniel Y. Fu and Stefano Ermon and Atri Rudra and Christopher R\'{e}",
  title     = "FlashAttention: Fast and Memory-Efficient Exact Attention with IO-Awareness",
  booktitle = "Advances in Neural Information Processing Systems (NeurIPS)",
  year      = "2022"
}

@inproceedings{fa2,
  author    = "Tri Dao",
  title     = "FlashAttention-2: Faster Attention with Better Parallelism and Work Partitioning",
  booktitle = "International Conference on Learning Representations (ICLR)",
  year      = "2024"
}

@inproceedings{fa3,
  author    = "Jay Shah and Ganesh Bikshandi and Ying Zhang and Vijay Thakkar and Pradeep Ramani and Tri Dao",
  title     = "FlashAttention-3: Fast and Accurate Attention with Asynchrony and Low-Precision",
  booktitle = "Advances in Neural Information Processing Systems (NeurIPS)",
  year      = "2024"
}

@inproceedings{fa4,
  author    = "Ted Zadouri and Markus Hoehnerbach and Jay Shah and Timmy Liu and Vijay Thakkar and Tri Dao",
  title     = "FlashAttention-4: Algorithm and Kernel Pipelining Co-Design for Asymmetric Hardware Scaling",
  booktitle = "Proceedings of Machine Learning and Systems (MLSys)",
  year      = "2026"
}

@inproceedings{armfa,
  author    = "Xiao Fu and Weiling Yang and Dezun Dong and Xing Su",
  title     = "Optimizing Attention by Exploiting Data Reuse on ARM Multi-core CPUs",
  booktitle = "Proceedings of the 38th ACM International Conference on Supercomputing (ICS)",
  year      = "2024"
}

@inproceedings{armsme,
  author    = "Chencheng Deng and Weiling Yang and Jianbin Fang and Dezun Dong",
  title     = "Demystifying ARM SME to Optimize General Matrix Multiplications",
  booktitle = "2026 IEEE International Parallel and Distributed Processing Symposium (IPDPS)",
  year      = "2026"
}

@article{blackwellbench,
  author  = "Aaron Jarmusch and Sunita Chandrasekaran",
  title   = "Microbenchmarking NVIDIA's Blackwell Architecture: An In-depth Architectural Analysis",
  journal = "arXiv:2512.02189",
  year    = "2026"
}

@manual{h100arch,
  author = "{NVIDIA}",
  title  = "{NVIDIA H100 Tensor Core GPU Architecture}",
  year   = "2022"
}

@manual{cuda124,
  author = "{NVIDIA}",
  title  = "{CUDA C++ Programming Guide}, Version 12.4",
  year   = "2024"
}

@techreport{blackwellbrief,
  author      = "{NVIDIA}",
  title       = "{NVIDIA Blackwell Architecture Technical Brief}",
  institution = "{NVIDIA}",
  year        = "2024"
}

@inproceedings{sageattn,
  author    = "Jintao Zhang and Jia Wei and Pengle Zhang and Jun Zhu and Jianfei Chen",
  title     = "SageAttention: Accurate 8-Bit Attention for Plug-and-Play Inference Acceleration",
  booktitle = "International Conference on Learning Representations (ICLR)",
  year      = "2025"
}

@inproceedings{sageattn2,
  author    = "Jintao Zhang and Haofeng Huang and Pengle Zhang and Jia Wei and Jun Zhu and Jianfei Chen",
  title     = "SageAttention2: Efficient Attention with Thorough Outlier Smoothing and Per-Thread INT4 Quantization",
  booktitle = "International Conference on Machine Learning (ICML)",
  year      = "2025"
}

@inproceedings{sageattn3,
  author    = "Jintao Zhang and Jia Wei and Pengle Zhang and Xiaoming Xu and Haofeng Huang and Haoxu Wang and Kai Jiang and Jun Zhu and Jianfei Chen",
  title     = "SageAttention3: Microscaling FP4 Attention for Inference and an Exploration of 8-Bit Training",
  booktitle = "Advances in Neural Information Processing Systems (NeurIPS)",
  year      = "2025"
}

@article{hifa4,
  author  = "Hui Dong and Yanzhao Li and Jie Gao and Chunlu Li and Zhiyuan Zhang and Yupeng Sun and Zhenyuan Chen and Zhiqiang Zou",
  title   = "HiFA4: Training-Free 4-bit FlashAttention on Ascend HIF4 NPUs for LLM Inference",
  journal = "arXiv:2607.04302",
  year    = "2026"
}

@inproceedings{fp4ht,
  author    = "Vage Egiazarian and Roberto L. Castro and Denis Kuznedelev and Andrei Panferov and Eldar Kurtic and Shubhra Pandit and Alexandre Marques and Mark Kurtz and Saleh Ashkboos and Torsten Hoefler and Dan Alistarh",
  title     = "Bridging the Gap Between Promise and Performance for Microscaling FP4 Quantization",
  booktitle = "International Conference on Learning Representations (ICLR)",
  year      = "2026"
}

@inproceedings{smoothquant,
  author    = "Guangxuan Xiao and Ji Lin and Mickael Seznec and Hao Wu and Julien Demouth and Song Han",
  title     = "SmoothQuant: Accurate and Efficient Post-Training Quantization for Large Language Models",
  booktitle = "International Conference on Machine Learning (ICML)",
  year      = "2023"
}

@inproceedings{gptq,
  author    = "Elias Frantar and Saleh Ashkboos and Torsten Hoefler and Dan Alistarh",
  title     = "GPTQ: Accurate Post-Training Quantization for Generative Pre-trained Transformers",
  booktitle = "International Conference on Learning Representations (ICLR)",
  year      = "2023"
}

@inproceedings{awq,
  author    = "Ji Lin and Jiaming Tang and Haotian Tang and Shang Yang and Wei-Ming Chen and Wei-Chen Wang and Guangxuan Xiao and Xingyu Dang and Chuang Gan and Song Han",
  title     = "AWQ: Activation-aware Weight Quantization for On-Device LLM Compression and Acceleration",
  booktitle = "Proceedings of Machine Learning and Systems (MLSys)",
  year      = "2024"
}

@inproceedings{kivi,
  author    = "Zirui Liu and Jiayi Yuan and Hongye Jin and Shaochen Zhong and Zhaozhuo Xu and Vladimir Braverman and Beidi Chen and Xia Hu",
  title     = "KIVI: A Tuning-Free Asymmetric 2bit Quantization for KV Cache",
  booktitle = "International Conference on Machine Learning (ICML)",
  year      = "2024"
}

@inproceedings{sglang,
  author    = "Lianmin Zheng and Liangsheng Yin and Zhiqiang Xie and Chuyue Sun and Jeff Huang and Cody Hao Yu and Shiyi Cao and Christos Kozyrakis and Ion Stoica and Joseph E. Gonzalez and Clark Barrett and Ying Sheng",
  title     = "SGLang: Efficient Execution of Structured Language Model Programs",
  booktitle = "Advances in Neural Information Processing Systems (NeurIPS)",
  year      = "2024"
}

@inproceedings{longbench,
  author    = "Yushi Bai and Xin Lv and Jiajie Zhang and Hongchang Lyu and Jiankai Tang and Zhidian Huang and Zhengxiao Du and Xiao Liu and Aohan Zeng and Lei Hou and Yuxiao Dong and Jie Tang and Juanzi Li",
  title     = "LongBench: A Bilingual, Multitask Benchmark for Long Context Understanding",
  booktitle = "Proceedings of the 62nd Annual Meeting of the Association for Computational Linguistics (ACL)",
  year      = "2024"
}

@inproceedings{ruler,
  author    = "Cheng-Ping Hsieh and Simeng Sun and Samuel Kriman and Shantanu Acharya and Dima Rekesh and Fei Jia and Yang Zhang and Boris Ginsburg",
  title     = "RULER: What's the Real Context Size of Your Long-Context Language Models?",
  booktitle = "Conference on Language Modeling (COLM)",
  year      = "2024"
}

@inproceedings{flashinfer,
  author    = "Zihao Ye and Lequn Chen and Ruihang Lai and Wuwei Lin and Yineng Zhang and Stephanie Wang and Tianqi Chen and Baris Kasikci and Vinod Grover and Arvind Krishnamurthy and Luis Ceze",
  title     = "FlashInfer: Efficient and Customizable Attention Engine for LLM Inference Serving",
  booktitle = "Proceedings of Machine Learning and Systems (MLSys)",
  year      = "2025"
}

@inproceedings{fasttree,
  author    = "Zaifeng Pan and Yitong Ding and Yue Guan and Zheng Wang and Zhongkai Yu and Xulong Tang and Yida Wang and Yufei Ding",
  title     = "FastTree: Optimizing Attention Kernel and Runtime for Tree-Structured LLM Inference",
  booktitle = "Proceedings of Machine Learning and Systems (MLSys)",
  year      = "2025"
}

@inproceedings{spatten,
  author    = "Hanrui Wang and Zhekai Zhang and Song Han",
  title     = "SpAtten: Efficient Sparse Attention Architecture with Cascade Token and Head Pruning",
  booktitle = "2021 IEEE International Symposium on High-Performance Computer Architecture (HPCA)",
  pages     = "97--110",
  year      = "2021"
}

@inproceedings{elsa,
  author    = "Tae Jun Ham and Yejin Lee and Seong Hoon Seo and Soosung Kim and Hyunji Choi and Sung Jun Jung and Jae W. Lee",
  title     = "ELSA: Hardware-Software Co-design for Efficient, Lightweight Self-Attention Mechanism in Neural Networks",
  booktitle = "2021 ACM/IEEE 48th Annual International Symposium on Computer Architecture (ISCA)",
  pages     = "692--705",
  year      = "2021"
}

@inproceedings{splitwise,
  author    = "Pratyush Patel and Esha Choukse and Chaojie Zhang and Aashaka Shah and \'{I}\~{n}igo Goiri and Saeed Maleki and Ricardo Bianchini",
  title     = "Splitwise: Efficient Generative LLM Inference Using Phase Splitting",
  booktitle = "51st Annual International Symposium on Computer Architecture (ISCA)",
  year      = "2024"
}

@inproceedings{podattn,
  author    = "Aditya K. Kamath and Ramya Prabhu and Jayashree Mohan and Simon Peter and Ramachandran Ramjee and Ashish Panwar",
  title     = "POD-Attention: Unlocking Full Prefill-Decode Overlap for Faster LLM Inference",
  booktitle = "Proceedings of the International Conference on Architectural Support for Programming Languages and Operating Systems (ASPLOS)",
  year      = "2025"
}

@inproceedings{mmlu,
  author    = "Dan Hendrycks and Collin Burns and Steven Basart and Andy Zou and Mantas Mazeika and Dawn Song and Jacob Steinhardt",
  title     = "Measuring Massive Multitask Language Understanding",
  booktitle = "International Conference on Learning Representations (ICLR)",
  year      = "2021"
}

@article{gms8k,
  author  = "Karl Cobbe and Vineet Kosaraju and Mohammad Bavarian and Mark Chen and Heewoo Jun and Lukasz Kaiser and Matthias Plappert and Jerry Tworek and Jacob Hilton and Reiichiro Nakano and Christopher Hesse and John Schulman",
  title   = "Training Verifiers to Solve Math Word Problems",
  journal = "arXiv:2110.14168",
  year    = "2021"
}

@inproceedings{HellaSwag,
  author    = "Rowan Zellers and Ari Holtzman and Yonatan Bisk and Ali Farhadi and Yejin Choi",
  title     = "HellaSwag: Can a Machine Really Finish Your Sentence?",
  booktitle = "Proceedings of the 57th Annual Meeting of the Association for Computational Linguistics (ACL)",
  year      = "2019"
}

@inproceedings{WinoGrande,
  author    = "Keisuke Sakaguchi and Ronan Le Bras and Chandra Bhagavatula and Yejin Choi",
  title     = "WinoGrande: An Adversarial Winograd Schema Challenge at Scale",
  booktitle = "Proceedings of the AAAI Conference on Artificial Intelligence (AAAI)",
  year      = "2020"
}

\end{document}